\documentstyle[preprint,aps]{revtex}
\tightenlines
\input{psfig.sty}
\begin{document}
\preprint{UNDPDK-98-04c}
\title{What the Atmospheric Neutrino Anomaly is Not}
\author{J.M.~LoSecco}
\address{University of Notre Dame, Notre Dame, Indiana 46556}
\date{\today}
\maketitle
\begin{abstract}

The atmospheric neutrino anomaly is the apparent reduction of the
$\nu_{\mu}/\nu_{e}$ ratio observed in underground detectors.
It represents either a reduction in the
muon neutrino interaction rate or an excess of the electron neutrino
interaction rate, or both.

Unable to answer the question of ``What else could it be?''
this paper explores a number of alternatives which do not seem to be viable.
Various methods to reduce the apparent muon rate or to increase the apparent
electron rate are discussed.

Perhaps our bias that the interactions are due to neutrinos of atmospheric
origin is incorrect.  Both of these assumptions need to be confirmed.
Efforts to reduce uncertainties in the estimated atmospheric neutrino
flux would also help to narrow the possibilities further.

Subject headings: Cosmic Rays --- Elementary Particles --- Neutrino Oscillations
\\
\end{abstract}


\section{Introduction}
The atmospheric neutrino anomaly \cite{haines,kamioka,imbo,suposc,kajita}
is the discrepancy
between the observed and expected rate of electron and muon neutrino
interactions in underground detectors.  In general it is believed that
these neutrinos originate in the Earth's atmosphere as a consequence of
the decay of short lived particles created by cosmic ray interactions.

The atmospheric neutrino anomaly is characterized by a low fraction of
muon neutrino interactions relative to the observed rate of electron neutrino
interactions \cite{supcont}.  Only 61\% of the expected rate is found.
The deficiency is energy independent in the
range 200 MeV to 4000 MeV.  The deficiency appears to be isotropic up to
at least 1200 MeV.
In particular neutrinos entering the detector from above
evidence the same muon neutrino deficiency as in other directions.

The distance scales in the problem span about three decades.  Neutrinos
coming from below have traveled about $1.3 \times 10^{7}$ meters.  Those
coming from above have traveled as little as $1.5 \times 10^{4}$ meters.

The uniformity of the effect in both direction and energy places very
strong constraints on possible physical explanations.

While there is strong agreement of the characteristics of the effect for
energies below about 1.2 GeV, IMB \cite{clark97} has failed to confirm the
multi-GeV effect reported by Kamioka \cite{multigev} and
Super Kamioka \cite{multig}.

The neutrinos in question have energies in the range of 200 MeV to a few
GeV.  Such neutrinos have been extensively produced and studied in
accelerator research programs over the past 30 years.  It is unlikely that
some detail of the neutrino interaction itself could be responsible.

Since the absolute neutrino flux estimate is uncertain by about 20\% it is
unclear if the anomaly reflects a 39\% attenuation of the muon neutrino
flux or a comparable excess of the electron neutrino flux.

In an attempt to reduce the effect of the absolute neutrino flux uncertainty
the quantity $R$ has been introduced.
\[
R = \frac{(\mu / e )_{DATA}}{(\mu / e )_{MC}} = 0.61 \pm 0.03 \pm 0.05
\]
$R$ removes the absolute normalization error by comparing the relative rates
of muon type and electron type interactions to the simulated, or expected
value of this ratio.

The 61\% rate reduction noted above is in fact about equally distributed
between a reduction in the number of muon type interactions and an
excess in the number of electron type interactions.  But given the
uncertainty in the normalization neither of these deviations are significant
in themselves.

\begin{figure}
\psfig{figure=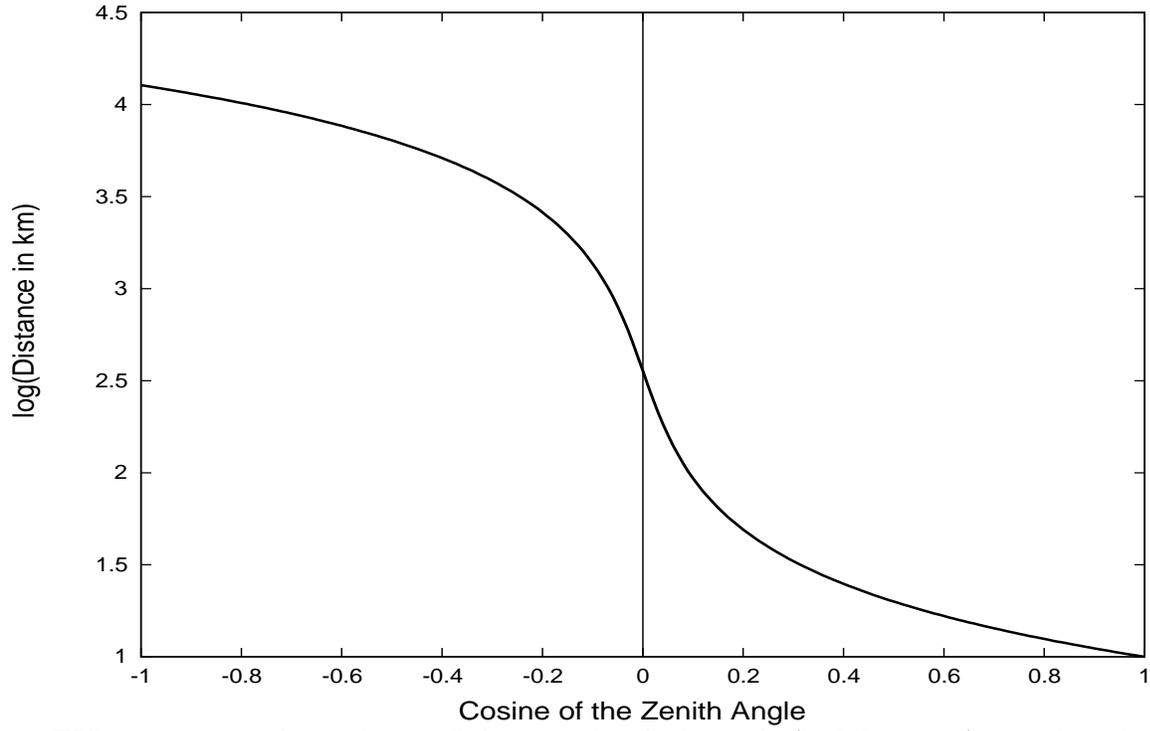,width=6.0in,height=3.8in}
\caption{\label{loglen} Approximate $\log_{10}$ of the neutrino flight path
(in kilometers) as a function of the cosine of the zenith angle.}
\end{figure}

The flight distances for the neutrinos from production to interaction vary
with direction.  Those coming from directly above have flight lengths on the
order of $1.5 \times 10^{4}$ meters.  Those coming from below travel as
much as $1.3 \times 10^{7}$ meters.  
Figure \ref{loglen} illustrates this range.

The suggestions offered here fall into roughly three categories.  Some
try to lower the apparent muon neutrino flux; some try to raise the
apparent electron neutrino rate and some try to do both.

\section{Neutrino Oscillations}
The interpretation of the effect in terms of neutrino oscillations seems to
run into difficulty with other experiments.  The small value of the observed
$\nu_{\mu}$ rate relative to the $\nu_{e}$ rate, even at
the short oscillation lengths available for the downward going flux pushes
the mass scale $\Delta m^{2}$ up \cite{problem} to regions ruled out by
various long baseline experiments.  The energy independence of the
effect \cite{bounds} also suggests a similar conclusion.
Mass scales on the order of 0.1-0.2 eV$^{2}$ seem to be needed.

\begin{figure}
\psfig{figure=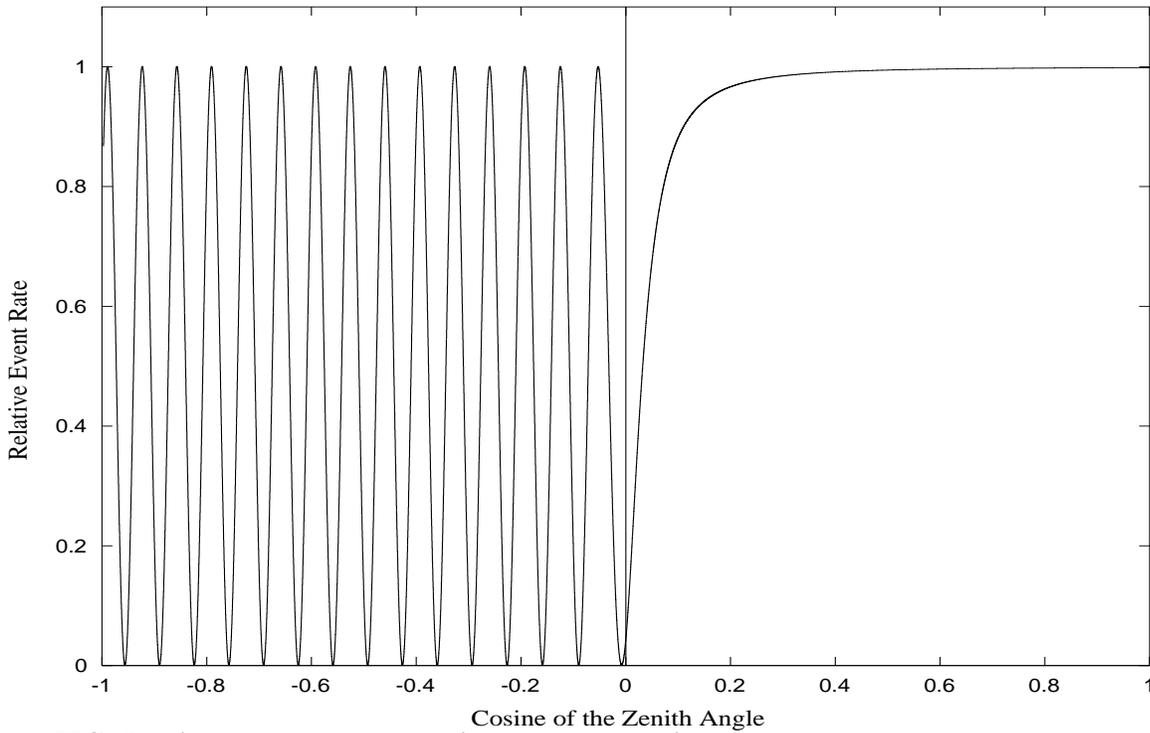,width=6.0in,height=3.8in,angle=270}
\caption{\label{nuosc} Angular distribution of $\nu_{\mu}$ interactions
for the suggested $\nu_{\mu} \rightarrow \nu_{\tau}$ neutrino
oscillation solution.}
\end{figure}

Figure \ref{nuosc} illustrates the expected angular distribution of
muon type events using the neutrino oscillation parameters of reference
\cite{suposc}, $\sin^{2}(2 \theta)=1$ and $\Delta m^{2} = 2.2 \times 10^{-3}$
eV$^{2}$.  The rapid oscillations in the lower hemisphere would be averaged
out to yield a net loss of half the rate in that region.

The anomaly is not subtle; it is a large effect.  If it is due to neutrino
oscillations it would require large mixing angles which, in general, are
easy to observe.  IMB failed to find evidence for $\nu_{\mu} \rightarrow
\nu_{\tau}$ oscillations \cite{svob} in a large sample of upward muons,
using two different methods.  The CHOOZ group \cite{chooz} has ruled out
$\overline{\nu_{e}} \rightarrow \overline{\nu_{x}}$ down to $\Delta m^{2}$
of $9 \times 10^{-4}$ eV$^{2}$.

\section{Neutrino Decay}
The effects observed might be produced via neutrino decay.  In particular
the decay $\nu_{\mu} \rightarrow \nu_{e} X$ would shift neutrinos from the
$\mu$ type to the $e$ type.  Decays of the form
$\nu_{\mu} \rightarrow \nu_{sterile} X$ would attenuate the $\nu_{\mu}$
flux.

For an initial number of neutrinos $N_{0}$ one finds after traveling a
distance $L$ with momentum $p$
\[
N=N_{0} e^{-\frac{m}{\tau} \frac{L}{p}}
\]
which is a function of $\frac{m}{\tau}$.  The dependence on $\frac{L}{p}$ is
substantially different from that of neutrino oscillations.

The limits \cite{pdg98} on $\frac{\tau}{m}$ are $> 15.4$ sec/eV for
$\nu_{\mu}$ and $> 300$
sec/eV for $\nu_{e}$.  With a maximal path length of about $1.3 \times 10^{9}$
cm (about $\frac{1}{23}$ of a second) and a minimum momentum of about
200 MeV/c,  $\frac{L}{p} <2.2 \times 10^{-10}$ sec/eV.
The maximum attenuation due to neutrino decay would be
$e^{-1.4 \times 10^{-11}}$ which is negligible.

\begin{figure}
\psfig{figure=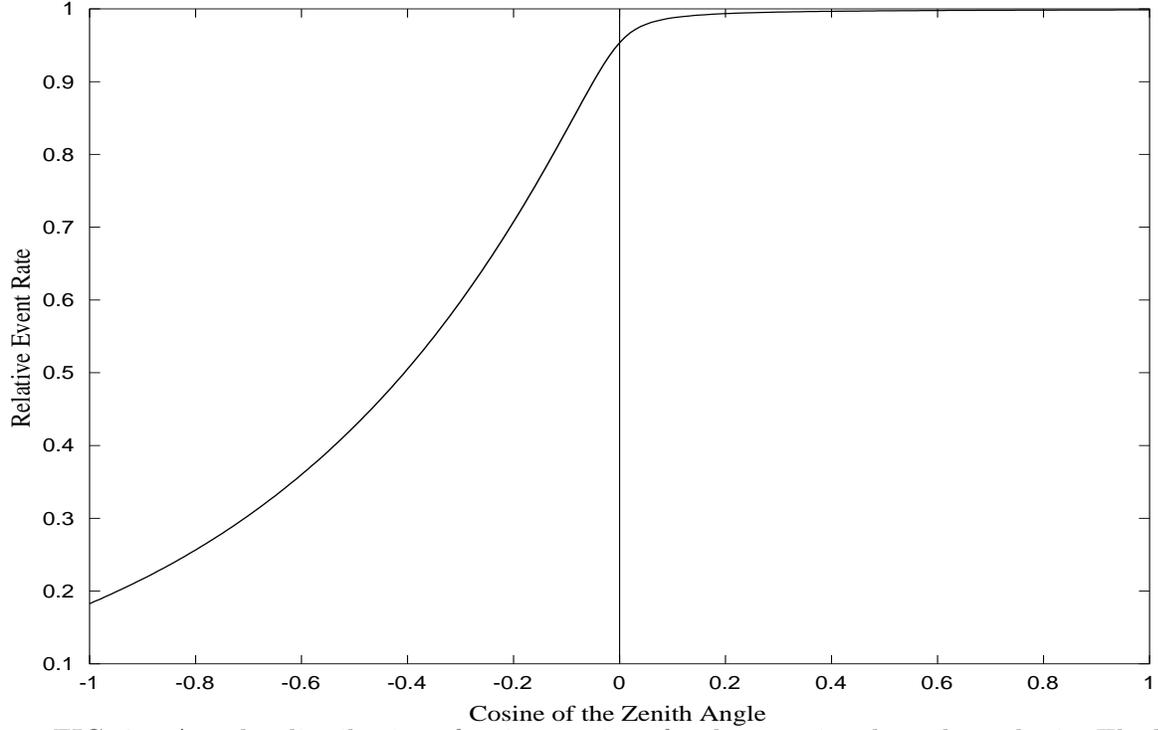,width=6.0in,height=3.8in,angle=270}
\caption{\label{nurot} Angular distribution of $\nu_{\mu}$ interactions
for the neutrino decay hypothesis.  The lifetime was taken as
$\frac{\tau}{m} = 0.033$ sec/GeV to enhance the effect.
The momentum is taken as 750 MeV/c.}
\end{figure}

Figure \ref{nurot} is an illustration of what neutrino decay might have
looked like.
In this figure $\frac{\tau}{m}$ was taken as  0.033 sec/GeV to enhance the
effect and show the consequences of the two distance scales.  The effects
of neutrino decay are only manifest in the lower hemisphere.

\section{Off Diagonal Neutral Currents}
Wolfenstein \cite{wolf78} has suggested that the weak neutral current might not
be flavor diagonal.  The neutrino part of the neutral current
could be of the form
\[
L = \cos^{2}a_{ab} (\overline{\nu_{a}}\nu_{a}+\overline{\nu_{b}}\nu_{b})
+ \sin^{2}a_{ab} (\overline{\nu_{a}}\nu_{b}+\overline{\nu_{a}}\nu_{b})
\]
Such an interaction would convert neutrinos of one flavor into another via
coherent forward scattering while moving through ordinary matter.
This would produce a transition probability that is energy independent.

On the other hand the transition probability would depend on the integrated
matter density along the path taken by the neutrinos.  Due to the inhomogeneous
composition of the Earth and the varying path lengths, the neutrino
composition would not be isotropic \cite{offdiag}.  Limits have been set
at $\sin^{2}a_{\mu\tau} < 0.1$.  The much larger data samples now available
and more refined methods of particle identification could be used to
improve the limit.

\begin{figure}
\psfig{figure=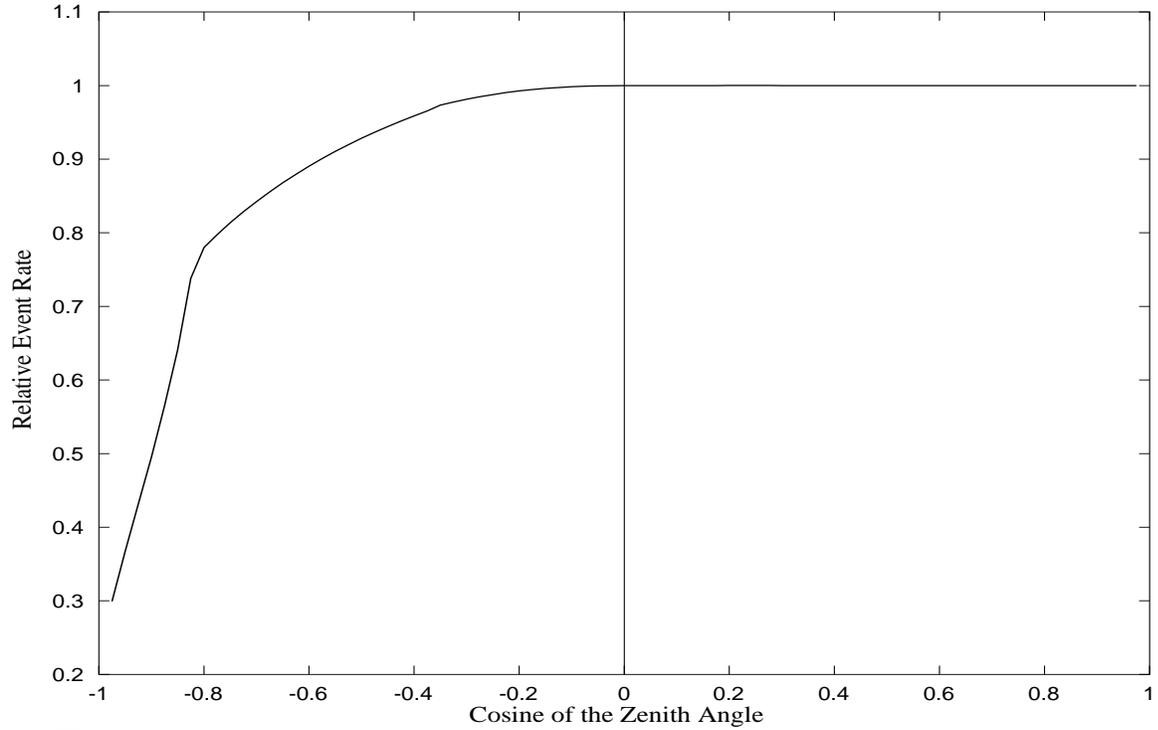,width=6.0in,height=3.8in,angle=270}
\caption{\label{nudiag} Angular distribution of $\nu_{\mu}$ interactions
for the off diagonal neutral current hypothesis.  The plot has been made
assuming that $\sin^2 (a_{\mu \tau}) = 0.1$.}
\end{figure}

Figure \ref{nudiag} illustrates this effect for $\sin^{2}a_{\mu\tau} = 0.1$.
The absence of much material in the upper hemisphere means that neutrinos
coming from above would not develop significant amplitude as $\nu_{\tau}$
so the downward component would be unchanged.  Some of the structure present
in figure \ref{nudiag} reflects details of the matter distribution in the
Earth's interior.

The absence of any significant reduction in the downward component of the
flux means that this method can not correctly reproduce the observed isotropy.

\section{Proton Decay}
It has been suggested \cite{manpdk} that the anomaly is attributable to
an excess of electrons due to the decay $p \rightarrow e \nu \overline{\nu}$.
Three body
proton decay would not produce any spectra above about 400-500 MeV.  But the
observed effect extends well above this energy.

\section{The Neutrino and Antineutrino Content}
If the source of neutrinos in the atmosphere were purely from the decay
of $\pi^{+}$ and its $\mu^{+}$ only $\nu_{e}$ and not $\overline{\nu_{e}}$
would be present in the atmospheric flux.  Indeed the ratio of
$\nu_{\mu}:\overline{\nu_{\mu}}:\nu_{e}:\overline{\nu_{e}}$ would be
approximately $1:1:1:0$ but the observed interaction rate would be much
closer to $1:0.33:1:0$\cite{pdg92} due to cross section differences.
So the observed ratio of muon like to electron like interactions would
be about 1.33.

If the simulations are done with a comparable rate of $\pi^{+}$ and $\pi^{-}$
the ratio of $\nu_{\mu}:\overline{\nu_{\mu}}:\nu_{e}:\overline{\nu_{e}}$
would be $2:2:1:1$ yielding an expected  ratio of muon like to electron like
interactions closer to $2:1$.  The observed over expected rate would then
be $1.33/2 = 0.67$ which is close to the reported value.  Careful
calculation of the neutrino energy spectrum could bring it down further.

If only $\pi^{-}$ were produced the flux ratio of
$\nu_{\mu}:\overline{\nu_{\mu}}:\nu_{e}:\overline{\nu_{e}}$ would be close
to $1:1:0:1$
leading to an interaction ratio of $1:0.33:0:0.33$ which would lead to a value
for the observed over expected rate of $\frac{1.33}{0.33} / \frac{2}{1}$
which is 2.

The detectors are not capable of distinguishing
an $e^{+}$ from an $e^{-}$.  It is unlikely that any electron detector based
on showers would be capable of making this distinction.  Indeed magnetic
measurements could distinguish an $e^{+}$ from an $e^{-}$ but magnetic
measurement is unrealistic for detectors of such large mass.

In a recent note\cite{losmu+-} it was pointed out that these detectors do
have the ability to distinguish $\mu^{+}$ from $\mu^{-}$ on a statistical
basis by exploiting the lifetime difference of the two particles in matter.

The observed
muon flux at ground level seems to be composed of an about 1.2:1 ratio of
$\mu^{+}$ to $\mu^{-}$.  But these do come from higher energies since the
muons of interest for the atmospheric neutrino flux would have already
decayed.  The primary interactions that produce the atmospheric neutrino
flux are comparable to those used to produce neutrinos at accelerator
laboratories, which have been able to produce effective research programs
with both $\nu_{\mu}$ and $\overline{\nu_{\mu}}$ beams.

The best published \cite{supcont} argument against the hypothesis of this
section is the decay rate
reported by underground experiments for the muon candidate events.  The
fraction of muon decays observed in this sample is as expected.  They see
67.6$\pm$1.6\% muon decays in the nonshowering sample.  The efficiency
for observing a $\mu^{+}$ decay is 80\% and the efficiency for observing
a $\mu^{-}$ decay is 63\%.  If the observed decay rate is a weighted
average of these two efficiencies, within the muon sample the flux ratio
of $\overline{\nu_{\mu}}$ to $\nu_{\mu}$ is about
1.1$\pm$0.5.  This assumes that the cross section ratio is $1:3$.
This is the expected flux ratio \cite{honda} but barely two sigma away from
zero.  This observation, while interesting, does not directly address the
primary
point of this section which is the $\nu_{e}$ and the $\overline{\nu_{e}}$
content of the underground flux.

There is a dearth of literature on $\overline{\nu_{\mu}}$ interactions
below 1 GeV and almost none for $\nu_{e}$ and $\overline{\nu_{e}}$ in the
energy range of most interest for the atmospheric neutrino anomaly.

Calculations \cite{honda} indicate that the atmospheric fluxes of
both neutrinos and antineutrinos are comparable to each other even at low
energies.  This is true for both muon and electron type neutrinos.
This is a bit puzzling since $\pi^{-}$ production in the atmosphere drops
faster than $\pi^{+}$ production as the primary cosmic ray energy is decreased.
For example, in 12.5 GeV/c proton collisions on Beryllium the
$\pi^{+}$ flux is about 3 times the $\pi^{-}$ flux \cite{zgs}.  For 3.7 GeV/c
proton proton collisions the $\pi^{+}$ seems to be produced about 5 times
as much as the $\pi^{-}$ \cite{hera}.  One might expect this production
asymmetry to be reflected in the modest energy neutrino fluxes.

Problems in estimating the low energy $\nu_{e}:\overline{\nu_{e}}$
flux ratio could not be extended to higher energy where production is
close to symmetric.

\section{Neutron Decay}
The decay $n \rightarrow p e^{-} \overline{\nu_{e}}$ produces electron
antineutrinos but no muon neutrinos.  One would expect copious production of
neutrons by cosmic ray interactions in the atmosphere.  The decay of these
neutrons would significantly raise the flux of electron neutrinos relative to
muon neutrinos.  The long flight paths possible in the atmosphere would permit
some of the neutrons to decay.

Unfortunately, for neutrons the $c \tau$ is about $2.7 \times 10^{8}$ km so
it is very unlikely that a significant number of them will have decayed before
being absorbed.  The permitted decay length in the atmosphere
is on the order of 20 km.
Only very slow neutrons would have the possibility of decaying and these could
not populate the high energy neutrino spectrum.

The decay length for extraterrestrially produced neutrons could be much
greater.  Neutron decay, or other phase space limited weak decays, might
contribute to a source which is not of atmospheric origin.

\section{Intermediate Lifetime Neutral Decay}
Is it possible that an as yet unknown neutral hadron, similar to the
neutron but with a much shorter lifetime could be decaying and produce
an excess $\nu_{e}$ flux?  Such an excess $\nu_{e}$ flux might not have been
seen at accelerators since the permitted decay times for producing
accelerator neutrinos are short.  Even the muons do not have time to decay
before they are absorbed or range out.

This might have been missed in accelerator neutrinos since the strongly
interacting
neutral would have been absorbed before it could decay.  But production
experiments with strong interactions make it unlikely that such an object
could have been missed.  Even if 100\% of it decays in the atmosphere
a flux comparable to that of the produced charged pions would be needed.

\section{Neutrino Production in Heavy Ion Collisions}
A significant fraction of the cosmic ray flux is in the form of ions heavier
than protons.  Is it possible that these are capable of producing a electron
neutrino rich high energy neutrino flux?

Fragmentation of the projectile ion could give the needed large boost.
Ions give one access to short lived beta unstable states.

It would seem to be very hard to get the rates needed.  The heavy ion modes of
production must compete with standard hadronic production via pions.
Pion production in heavy ion collisions is still very large so the impact on
the overall neutrino flux of weak decays of ions would be small if any.

To populate the neutrino spectrum in the energy range of about 1 GeV
would require heavy ions with a gamma factor on the order of 100.
Due to the steeply falling nature of the cosmic ray spectrum \cite{mont}
(approximately $E^{-3}$)
it seems difficult to get enough high energy heavy ion flux to compete
with the modest energy pion production of neutrinos.

\section{Violation of Muon Neutrino Electron Neutrino Universality}
If the electron neutrino cross section were significantly greater than
that of the muon neutrino the observed rates would not reflect the
true relative fluxes.  They would be weighted by the relative cross sections.
The atmospheric observations represent a very large sample of
$\nu_{e}$ interactions, perhaps even larger than accelerator samples.
Most neutrino observations in the energy range of
interest have been studied at accelerators using beams of $\nu_{\mu}$
and $\overline{\nu_{\mu}}$ so we have little direct evidence about
$\nu_{e}$ interactions at these energies.  But $\nu_{\mu}$ $\nu_{e}$
universality
seems to be well established in particle decays where phase space effects can
account for all differences in branching ratios.

In $\tau$ decay the radiative corrections can be calculated and the
ratio of muon decay to electron decay is expected to be 0.973.
Observations \cite{pdg98} find this to be 0.975$\pm$0.006.  Universality
can be violated by no more than 1\% which is insufficient to explain the
anomaly.

\section{Extraterrestrial Sources and Dark Matter}
In a prior publication \cite{losec1} the possibility that an excess neutral
flux of extraterrestrial origin, perhaps dark matter, was explored.
The decay of ambient dark matter can be ruled out since it would require
an energy density greater than the closure density of the universe.

An excess $\nu_{e}$ flux on the order of 0.4 neutrinos/cm$^{2}$/second
with a mean energy of about 500 MeV could explain the anomaly.  But there is
no obvious astrophysical origin for such $\nu_{e}$'s.  In general sources
that would produce $\nu_{e}$ of this energy are also quite capable of producing
$\nu_{\mu}$.  Even neutron decay, which was considered above, would be
associated with high energy collisions that would also produce $\nu_{\mu}$.
Assigning the anomaly to extraterrestrial sources does not solve the problem.
It merely shifts it to a different venue where more production methods
may become possible.

\section{Systematic Errors}
A systematic error that misclassified events as electrons when they came
from a muon neutrino interaction could explain the effect.  This seems
unlikely.  The effect was first seen as a deficiency of muon
decays by IMB \cite{haines}.  This method is independent of the particle
identification methods based on showering and nonshowering tracks.  The muon
decay method can be checked against stopping muons and against
the nonshowering data sample.  The
two methods confirm each other and have been checked against each other
\cite{imbo}.

Initially most iron calorimeter detectors \cite{frej} failed to see
the anomaly, which implied a possible systematic problem with water detectors.
But now the Soudan-2 detector \cite{allison} seems to confirm the existence
of an anomaly.


\section{Neutral Currents}
Neutral current interactions of $\nu_{\mu}$ would, in general go undetected
in these large devices or would be classified as showering types of events
and hence would more often
be attributed to $\nu_{e}$.  But the simulations have included the
neutral current effects \cite{supcont} and so this systematic effect
should be reflected
in the expected values.

\section{Conclusions}
The obvious question about the atmospheric neutrino anomaly is how do we know
that the interactions are due to neutrinos?  Even if they are due to neutrinos
how do we know that all of the neutrinos observed in underground detectors
are of atmospheric origin?

Indeed the total rate of observed interactions agrees well with the predicted
rates.  So the only opportunity for additional interactions, or attenuations
must be hidden in the 20\% uncertainty in the absolute flux
normalization \cite{honda}.

If what is being observed is not of atmospheric origin why does it
track the atmospheric spectrum so well?  The physics of neutrino production
in low density materials is universal so one might anticipate a similar
spectrum for cosmic ray induced neutrinos that are not of atmospheric
origin.  So why should such a source be $\nu_{\mu}$ deficient or $\nu_{e}$
enhanced?

It may be very difficult to measure the nonatmospheric contribution to the
ambient flux especially if it lacks any directional modulation.  It does not
seem feasible to do such large scale experiments outside of the Earth's
polluting atmosphere for the foreseeable future.

The most productive direction at present would be to reduce the uncertainty
on the absolute flux normalization \cite{fogli}.  The high statistics 
neutrino data samples now available mean that the measurements are dominated
by this source of systematic error.  Studies of the primary and secondary
cosmic ray flux should carefully quantify the relative production of
$\pi^{+}$ and $\pi^{-}$ either directly or via the high altitude muon
spectrum.

It is worth quoting the initial perspective on this effect,
``This discrepancy could be a statistical fluctuation or a systematic error
due to {\em (i)} an incorrect assumption as to the ratio of muon $\nu$'s to
electron $\nu$'s in the atmospheric fluxes, {\em (ii)} an incorrect estimate
of the efficiency for our observing a muon decay, or {\em (iii)} some other
as-yet-unaccounted-for physics'' \cite{haines}
For the most part these words are still true today.
\section*{Acknowledgments}
I would like to thank Jim Wilson and John Poirier for the opportunity to
discuss some of these ideas.  Neal Cason helped me track down the data
on inclusive pion production.  I would like to thank Bill Shephard for a
careful reading of the manuscript.

\end{document}